\documentstyle[amstex,amssymb,aps]{revtex}

\begin{document}
\title{Aspects of Duality in Nodal Liquids}
\author{Nick E. Mavromatos and Sarben Sarkar}
\address{Department of Physics, Theoretical Physics \\
King's College London\\
Strand, London WC2R 2LS}
\maketitle

\begin{abstract}
Starting from a microscopic $t-J$ like model and a $SU\left( 2\right) $
spin-charge separation ansatz, a relativistic continuum gauge lagrangian is
obtained in the vicinity of a nodal point of the Fermi surface. The
excitations in the pseudogap phase are described by topological excitations
in the dual model which has a $Z_{2}$ global symmetry due to the effect of
instantons. Confinement of spinon and holons emerge from this picture. The
adjoint and fundamental strings are associated with stripes. As the spin gap
decreases a local $Z_{2}$ symmetry emerges.\newline
\end{abstract}


\begin{centering} 

P.A.C.S. numbers: 11.15.Me,74.20.Mn 

\end{centering} 

\section{Introduction}

{\rm High temperature superconductors (HTS) are materials which are believed
by many to involve strong electron correlations \cite{And} and so they are
not in the universality class of BCS superconductors. Some believe that new
concepts familiar in field theory such as strings are relevant to their
description. The phase diagram for HTS is complex involving
antiferromagnetism, non-Fermi liquid behaviour, pseudogaps etc.\cite{Var}
However, partly because the models are not solvable the basic mechanisms are
still controversial. At the same time it is believed (or hoped) that the
gross features should have some underlying explanation based, not on the
microscopic details, but on simple general principles which underpin the
physics of all parts of the phase diagram. We shall present a model which
has a plausible justification from one of the generic models that are in
favour with the practitioners of high }$T_{c}${\rm \ and is based on the
inspired surmise that spin and charge somehow separate. This is familiar in
condensed matter from studies of exactly solvable, typically lattice, models
in one dimension. There is a sizeable and beautiful literature on this which
comes under the heading of the Bethe ansatz, since in the distant past Bethe
solved the one dimensional quantum Heisenberg model by ingenious methods
that bear his name.With the passage of time the deep reason for Bethe's
success was uncovered and this was linked to an infinite number of
conservation laws in such models. The bold extension of these ideas to two
dimensions has been made, reinforced by the presence of
quasi-one-dimensional structures that will be mentioned later. It is clear
that the formal and rigorous grounds for Bethe's success do not hold in two
dimensions; so we need to have a different route to the understanding of
excitations in HTS. The spin-charge separation that we will use does not
directly spring from the Bethe ansatz result concerning the true excitations
in one-dimensional systems. Rather it is a basis in which to define a model%
\cite{Ander}. The originators had hoped that the basis would be a good
starting point for the study of lattice models based on the experiences in
one dimension. Irrespective of the pros and cons of such an approach we will
employ it to derive a starting hamiltonian from a lattice }$t-J$ model. ( $t$
denotes the magnitude of the hopping matrix element and $J$ denotes the
Heisenberg antiferromagnetic exchange.) Just as in the non-linear $\sigma $
model, there are constraints to make the dynamics more strongly coupled. The
lattice constraint requires at most one hole on a site.{\rm \ It has been
argued that a simple one-band model (i.e. one involving one type of charged
carrier `orbital') may emerge as an approximation from a more realistic
model involving Cu} $3d_{x^{2}-y\text{ }}${\rm orbital} {\rm and the O }$%
\left( 2p_{x},2p_{y}\right) $ {\rm orbitals \cite{tj}. Even if the holes are
localised on the O atoms, the lowest energy states can be (Zhang-Rice)
singlets formed with the hole on the Cu atom which can hop as a fermionic
entity from plaquette\ to plaquette formed by nearest neighbour O atoms~\cite%
{Feiner}. This is not a rigorous picture though, and features such as direct
hopping of O holes, of magnitude }$t_{p}$,{\rm \ is ignored ( as opposed to
that induced by hybridization of the O and Cu holes).} The details are
involved. Moreover the bare values are not necessarily the ones that are
pertinent to the low energy physics (in the sense of running couplings used
for the renormalisation group); we shall, as is traditional, ignore such
complications and consider just the Zhang-Rice singlet picture. Given this
input and the spin-charge separation ansatz, a certain amount of interesting
and non-trivial phenomenology will arise.{\rm \ Inherent to the spin-charge
ans\"{a}tze are constraints and hidden gauge symmetries \cite{Ander}\cite%
{Dor}\cite{Far}. The exploitation of this gauge symmetry, using duality
arguments developed for the description of confinement in }$2+1$ {\rm gauge
symmetries, will form the central plank of our approach to stripe phases %
\cite{stripes} and the resonant valence bond (RVB) states of high }$T_{c}$%
{\rm \ materials. In the dual description we shall see that, in general, a }$%
Z_{2}$ {\rm gauge theory will emerge (but from a perspective quite different
to that used in some recent work \cite{zaanen}). }

{\rm \ Although various widely accepted effective lattice hamiltonians are
simple to write down, they cannot be solved rigorously for planar systems.
Consequently phenomenological effective continuum low energy hamiltonians,
which incorporate important symmetries, have been a focus of attention \cite%
{Balents} . A spur to this has been recent experimental progress in
establishing properties of the pseudogap phase in the underdoped HTS \cite%
{Campuzano}. Whereas the superconducting phase has a well defined Fermi
surface and quasiparticles, the underdoped region has a Fermi surface but
not well defined quasiparticles. Moreover the pseudogap \cite{Stanf} itself
has the }$d${\rm -wave symmetry of the superconducting gap. There is some
smoothness in the transition from the superconductor to the pseudogap. In
the light of this, the low energy properties of the theory in the
neighbourhood of the zeros or nodes of the superconducting gap may be a good
starting point for investigating the physics of the pseudogap phase\cite%
{Balents}. Near the nodes the theory can be described by a relativistic
theory\cite{Far}.This is an example of the intimate relationship between
different parts of the phase diagram. Moreover the inhomegeneities of the
charge distribution inherent in the formation of stripes gives a physical
separation of charges and spins. Our approach, which is restricted to low
doping, will allow us to describe this separation in terms of the confining
properties of \ }$2+1$ {\rm dimensional spontaneously broken }$SU\left(
2\right) ${\rm \ gauge theories \cite{Mavro} in a dual theory for nodal
liquids. Moreover the existence of skyrmion textures centered on the doped
holes \cite{marinoskyr} in this picture support an antiphase boundary
condition for the staggered spins on either side of the stripe.}

{\rm Gauge theories have figured in theories of high temperature
superconductivity since the early days of the subject \cite{Ander}.
Furthermore the strong coupling Hubbard model at half-filling, which leads
to the Heisenberg model for a quantum antiferromagnet, was mapped onto a
lattice gauge theory with }$SU\left( 2\right) ${\rm \ gauge group and
Nambu-Dirac fermions \cite{Affleck}. Holes in antiferromagnets were studied
by Shankar \cite{Shan} in one-space dimensions in the limit of large spin
and small doping. This was generalised by Dorey and Mavromatos to two-space
dimensions \cite{Dor}. The large spin limit of the anti-ferromagnet is the
same limit used by Haldane in his derivation of the }$\sigma ${\rm (}$CP^{1}$%
{\rm )-model \cite{Hal}. Dorey and Mavromatos, and Shankar modelled the
structure of copper oxide layers in HTS using a bipartite lattice. In the\
large spin limit holes hop only within each sublattice. This is due to
assumed underlying Ne\'{e}l order and orthogonality properties of spin
coherent states. The labelling of sublattices provides a flavour label for
the fermions. The concepts of large spin and low doping will also be assumed
in our analysis. However properties of the low energy theory around the
nodes of the superconducting gap were not of interest at that time. It is
interesting to note that, quite recently, by investigating Marshall wave
functions for spin }$\frac{1}{2}${\rm \ Heisenberg antiferromagnets,
qualitatively similar Berry phases were found for a hole performing a closed
circuit \cite{Weng}. Indeed these authors argued that this phase was
fundamental to vanishing spectral weight and non-Fermi liquid behaviour.
These phases can be incorporated in a statistical }$U_{S}\left( 1\right) $
gauge factor in the hopping factor of the Zhang-Rice singlet.{\rm \ }

{\rm In this article we will show how, starting from a microscopic lattice
gauge theory, for which arguments have been given at length elsewhere \cite%
{Far}, some of the features of the pseudogap and stripes can emerge due to
general features of symmetry and confining properties of gauge theories in }$%
2+1$ {\rm dimensions. Our aim is to incorporate certain symmetries within a
spin-charge separation ansatz as a relativistic theory in the neighbourhood
of a node. The framework is a development of the earlier work of Dorey and
Mavromatos \cite{Dor} and Farakos and Mavromatos \cite{Far}. The structure
of the article is the following: in section II we review how one derives the
continuum field theory of the nodal excitation under the non-Abelian spin
charge separation ansatz in case of doped antiferromagnetic systems with an
approximate particle-hole symmetry. In section III we first derive in detail
the dual theory in a Landau-Ginzburg approach in the case where the spinon
degrees of freedom are assumed very massive so as to decouple. In section IV
we discuss domain wall structures in this approach, which we connect with
stripes in the original microscopic theory. In section V we incorporate the
spinon degrees of freedom and show how, as the spin gap decreases, a local $%
Z_{2}$ emerges in the theory. We also discuss antiphase properties of the
staggered magnetization in the microscopic picture (as opposed to the dual
one) by making the observation that the domain wall structure in the dual
Lagrangian corresponds to a domain wall structure in the flux of the
statistical photon, and stripe conductivity by discussing the Dirac equation
of the nodal holons in such flux backgrounds. Conclusions and outlook are
presented in section VI. }

\section{\rm From Microscopic models to Continuum Relativistic Field Theory}

{\rm It was noted by Affleck et al \cite{Affleck}\ that\ the spin operator
at site }$i${\rm ,\ in terms of the Pauli spin matrices }$\vec{\sigma}${\rm %
, can be written as \ }%
\begin{equation}
\vec{S}_{i}=Tr\left( \chi _{i}^{\dagger }\chi _{i}\vec{\sigma}\right)
\end{equation}

{\rm where \ }$\chi =\left( 
\begin{tabular}{ll}
$c_{1}$ & $c_{2}$ \\ 
$c_{2}^{\dagger }$ & $-c_{1}^{\dagger }$%
\end{tabular}%
\right) ${\rm \ and }$c_{1}${\rm \ and }$c_{2}${\rm \ are the annihilation
operators for up and down holes (which will actually represent the
Zhang-Rice singlets). We will make a spin-charge separation ansatz \cite{Far}%
\ which has slave fermion character and incorporates the }$CP^{1}${\rm \
constraint known from large-spin analysis at half-filling ( the Heisenberg
limit). Our ansatz is }%
\begin{equation}
\chi =\left( 
\begin{tabular}{ll}
$\psi _{1}$ & $\psi _{2}$ \\ 
$\psi _{2}^{\dagger }$ & $-\psi _{1}^{\dagger }$%
\end{tabular}%
\right) \left( 
\begin{tabular}{ll}
$z_{1}$ & $-\bar{z}_{2}$ \\ 
$z_{2}$ & $\bar{z}_{1}$%
\end{tabular}%
\right) =\Psi Z  \label{scan}
\end{equation}

{\rm \bigskip }

{\rm where }$\psi _{\alpha }${\rm \ and }$z_{\alpha }${\rm , }$\alpha =1,2$%
{\rm \ are fermions and bosons representing holes and spins repectively .
The }$\alpha ${\rm \ index labels the sublattice structure of the bipartite
lattice (relevant for the large spin analysis). The ansatz is valid in the
low doping situation \cite{Far} and on assuming canonical (anti-)commutation
relations for }$\left( \psi _{\alpha }\right) ${\rm \ }$z_{\alpha }${\rm ,
canonical anticommutation for }$c_{\alpha }${\rm \ hold provided}

\begin{equation}
{\rm \psi }_{1,i}{\rm \psi }_{2,i}{\rm =0=\psi }_{2,i}^{\dagger }{\rm \psi }%
_{1,i}^{\dagger }{\rm ,\ \ \ \ \ \ }\sum_{\beta =1}^{2}\left( \bar{z}_{\beta
,i}z_{\beta ,i}+\psi _{\beta ,i}^{\dagger }\psi _{\beta ,i}\right) {\rm =1.}
\end{equation}

{\rm \bigskip }

{\rm These are single occupancy constraints which implies that the ansatz
requires strong coupling. There is also a certain {\it redundancy} in the
description due to the following }$SU\left( 2\right) ${\rm \ gauge symmetry}

{\rm \ }%
\begin{equation}
\Psi _{i}\rightarrow \Psi _{i}h_{i},\ \ \ \ \ Z_{i}\rightarrow
h_{i}^{\dagger }Z_{i}\;\text{{\rm where}}\;h_{i}\in SU\left( 2\right) .
\end{equation}%
{\rm The global }$SU\left( 2\right) ${\rm \ spin manifests itself as }%
\begin{equation}
Z_{i}\rightarrow Z_{i}h  \label{ROTNL}
\end{equation}%
{\rm where }$h${\rm \ is a {\it global} transformation in }$SU\left(
2\right) ${\rm . However, there is also a dynamical $U_{s}(1)$ gauge
symmetry acting only on the $\Psi $ fields, which is due to phase
frustration from holes moving in a spin background. Some arguments to
justify this from a microscopic point of view have been given in reference~%
\cite{Dor}\cite{Weng}. Consequently, this symmetry is associated with exotic
statistics of the pertinent excitations~\cite{Far}, which is an exclusive
feature of the planar spatial geometry. }

{\rm By similar reasoning to Farakos and Mavromatos \cite{Far} we arrive at
a generalised Hartree-Fock hamiltonian }$H_{HF}$ of the holon-spinon
excitations:

\begin{equation}
\ H_{HF}=\sum_{\langle ij\rangle }Tr\left( 
\begin{array}{c}
A_{1}t_{ij}\Psi _{i}^{\dagger }\left( 1+\sigma _{3}\right) U_{ij}R_{ij}\Psi
_{j}+ \\ 
A_{2}Z_{i}^{\dagger }R_{ij}Z_{j}+h.c.%
\end{array}%
\right) .
\end{equation}%
{\rm where }$U_{ij\text{ }}${\rm (}$R_{ij}${\rm )} {\rm is the} $U_{s}\left(
1\right) $ {\rm (}$SU\left( 2\right) ${\rm ) link element} {\rm between
nearest neighbour sites }$i$ {\rm and} $j$ {\rm ( }$A_{1}$ {\rm and} $A_{2}$ 
{\rm are constants resulting from assumed frozen amplitudes assumed
typically in lattice gauge treatments). }

{\rm A necessary condition for a relativistic theory is that the fermions
are coupled to the }$U_{{\em s}}\left( 1\right) ${\rm \ gauge potential \
which fluctuates around a background with a flux of }$\pi ${\rm \ per
plaquette in each sublattice. The flux phase background \cite{flux},
implying that a product of background\ }$U_{{\em s}}\left( 1\right) ${\rm \
gauge potentials around each sublattice plaquette equals }$\left( -1\right) $%
{\rm , is essential for yielding relativistic fermions in the continuum
limit according to the general analysis of Burden and Burkitt \cite{Bur} .
Such considerations lead to the form of our effective starting lagrangian.
Instead of the }$2\times 2${\rm \ matrix structure of the fermion fields }$%
\Psi ${\rm \ it is convenient \cite{Far} to work in terms of two Nambu }$2$%
{\rm -component spinors:}%
\begin{equation}
\tilde{\Psi}_{1}^{\dagger }=\left( \psi _{1},\psi _{2}^{\dagger }\right)
\end{equation}

{\rm and }%
\begin{equation}
\ \tilde{\Psi}_{2}^{\dagger }=\left( \psi _{2},-\psi _{1}^{\dagger }\right) .
\end{equation}%
{\rm In the background field of the flux phase of the }$U_{{\em s}}\left(
1\right) ${\rm \ field a Kawamoto-Smit \bigskip transformation \cite{Smit}
to another set of }$2${\rm -component spinor }$\Xi _{\alpha }${\rm \ is
made. The components of }$\Xi ${\rm \ are linearly related to the components
of }$\Psi ${\rm . In terms of these new spinors we obtain, in the action for
the fermions,\ the following term which is necessary to have a continuum
limit with Dirac structure: }%
\begin{equation}
S_{F}=\frac{1}{2}\kappa ^{\prime }\sum_{i,\mu }\left( \bar{\Xi}_{i}\left(
-\gamma _{\mu }\right) R_{i,\mu }U_{i,\mu }\Xi _{i+\mu }+\bar{\Xi}_{i+\mu
}\gamma _{\mu }R_{i,\mu }^{\dagger }U_{i,\mu }^{\dagger }\Xi _{i}\right)
\end{equation}%
{\rm Here }$\bar{\Xi}${\rm \ denotes }$\Xi ^{\dagger }\gamma _{0}${\rm \ and
the representation of the Dirac matrices that we use is }$\gamma _{\mu
}=i\left( \sigma _{3},\sigma _{1},\sigma _{2}\right) ,\;\mu =0,1,2${\rm . }$%
U_{i,\mu }${\rm \ is the gauge link variable }$U_{i}\,_{i+\mu }${\rm . The
gauge invariant kinetic terms for the link variables are taken to be }%
\begin{equation}
S_{G}=\sum_{p}[\beta _{2}\left( 1-Tr\,R_{p}\right) +\beta _{1}\left(
1-Tr\,U_{p}\right) ]
\end{equation}%
{\rm where }$p${\rm \ denotes plaquettes, \ the }$\beta _{j}${\rm \ are the
inverse couplings; }$R_{p}${\rm \ and }$U_{p}${\rm \ symbolize the plaquette
product over the }$SU\left( 2\right) ${\rm \ and }$U_{{\em s}}\left(
1\right) ${\rm \ link variables repectively. The inverse coupling }$\beta
_{2}${\rm \ is assumed large in order to have a conventional continuum Dirac
form for the holon excitations. It should be noted that the kinetic terms
for the link variables appear in any case if the spinon degrees of freedom
are in the phase where they are massive (i.e pseudogap phase) due to quantum
fluctuations . In terms of the microscopic parameters of the underlying }$%
t-J ${\rm \ model, the }$SU\left( 2\right) ${\rm \ dimensionful coupling }$e$%
{\rm \ \ (}$\beta _{2}\varpropto \frac{1}{e^{2}}${\rm ) has a} spinon
contribution of the form $e^{2}\sim J\left( 1-\delta \right) $ where $\delta 
$ denotes doping concentration \cite{Dor}{\rm . Clearly the coupling due to
the spinon contribution decreases as the doping increases. This implies that
the gauge coupling for low doping is large. The kinetic term for the }$U_{%
{\em s}}\left( 1\right) ${\rm \ gauge is not generated by quantum
fluctuations of the spinons (which do not couple directly to the }$U_{{\em s}%
}\left( 1\right) ${\rm \ gauge potential). This is equivalent to the
condition }$\beta _{1}\thicksim 0.${\rm \ Since }$\beta _{1}=0${\rm , the }$%
U_{S}\left( 1\right) ${\rm \ gauge field on the lattice can be integrated
exactly \cite{McNeill}, in the path integral for the partition function, to
leave }%
\begin{equation}
\int {\cal D}{\it R\,}{\cal D\bar{\Xi}\,D\Xi \,}\exp \left( -S_{eff}\right)
\end{equation}%
{\rm where }%
\begin{equation}
S_{eff}=\beta _{2}\sum_{p}\left( 1-Tr\,R_{p}\right) +\sum_{i,\mu }\log
\,I_{0}\left( \sqrt{y_{i\mu }}\right)  \label{Efflattlag}
\end{equation}%
{\rm and }%
\begin{equation}
y_{i\mu }=-\kappa ^{2}Tr\left( M^{\left( i\right) }\left( -\gamma _{\mu
}\right) R_{i\mu }M^{\left( i+\mu \right) }\gamma _{\mu }R_{i\mu }^{\dagger
}\right) .
\end{equation}%
$\kappa ${\rm \ is related to }$\kappa ^{\prime }${\rm \ and is a constant
depending on the microscopic energy parameters of the underlying theory and
in particular the doping. }$M${\rm \ denotes a composite variable}$%
\;M_{ab,\alpha \beta }=\Xi _{b,\beta }\bar{\Xi}_{a,\alpha }${\rm , which can
be expanded in terms of a complete basis of bilinear fields \cite{Far}:}%
\begin{equation}
M^{\left( i\right) }=\sum_{a=1}^{3}\Phi _{a}\left( i\right) \sigma _{a}+%
{\cal S}_{4}\left( i\right) 1_{2}+i\left( \left( {\cal S}_{\mu }\right)
_{4}\left( i\right) \,\,\gamma ^{\mu }+\sum_{a=1}^{3}\left( \digamma _{\mu
}\right) _{a}\left( i\right) \;\gamma ^{\mu }\sigma _{a}\right) .
\end{equation}%
{\rm Here} 
\begin{equation}
\Phi _{^{1}}=-i\left( \bar{\Xi}_{1}\Xi _{2}-\bar{\Xi}_{2}\Xi _{1}\right)
,\;\Phi _{2}=\left( \bar{\Xi}_{1}\Xi _{2}+\bar{\Xi}_{2}\Xi _{1}\right)
,\,\Phi _{3}=\left( \bar{\Xi}_{1}\Xi _{1}-\bar{\Xi}_{2}\Xi _{2}\right)
\end{equation}%
{\rm form an adjoint representation of} $SU\left( 2\right) $, 
\begin{equation}
\left( {\cal A}_{\mu }\right) _{1}=i\left( \bar{\Xi}_{1}\tilde{\sigma}_{\mu
}\Xi _{2}-\bar{\Xi}_{2}\tilde{\sigma}_{\mu }\Xi _{1}\right) ,\;\left( {\cal A%
}_{\mu }\right) _{2}=\left( \bar{\Xi}_{1}\tilde{\sigma}_{\mu }\Xi _{2}+\bar{%
\Xi}_{2}\tilde{\sigma}_{\mu }\Xi _{1}\right) ,\;\left( {\cal A}_{\mu
}\right) _{3}=\left( \bar{\Xi}_{1}\tilde{\sigma}_{\mu }\Xi _{1}-\bar{\Xi}_{2}%
\tilde{\sigma}_{\mu }\Xi _{2}\right)
\end{equation}%
{\rm form a vector adjoint representation, the} ${\cal S}_{4}${\rm \ and }$%
\left( {\cal S}_{\mu }\right) $ {\rm are two singlets and }$\tilde{\sigma}%
_{0}=-i\sigma _{3},\tilde{\sigma}_{1}=\sigma _{1},\tilde{\sigma}_{2}=\sigma
_{2}.${\rm \ Since }$\Xi ${\rm \ is grassmanian, the Taylor series in }$\Xi $%
{\rm \ truncates and so }%
\begin{equation}
-\log I_{0}^{tr}\left( 2\sqrt{y_{i\mu }}\right) =-y_{i\mu }+\frac{1}{4}%
y_{i\mu }^{2}-\frac{1}{9}y_{i\mu }^{3}+\frac{11}{192}y_{i\mu }^{4}\ 
\end{equation}%
{\rm which is an exact expression. This effective potential allows us to
consider the dynamical formation of various condensates of fermionic
bilinears. It is also necessary to pass from the path integral over
fermionic variables }$\Xi ,\bar{\Xi}${\rm \ to the bosonic variables in
terms of which }$M$ is written {\rm . The transformation has a non-trivial
Jacobian \cite{McNeill}, which leads to additional terms in the effective
action of the form }$\ -\sum_{i,\mu }\frac{1}{6}\log \det M_{i}M_{i+\mu }$%
{\rm \ (where }$i${\rm \ is a lattice site index and }$\mu ${\rm \ is a
space-time index ). In the ground state a scalar condensate along the }$%
\sigma _{3}${\rm \ direction is formed, i.e.}%
\begin{equation}
\langle M^{\left( i\right) }\rangle =u\sigma _{3}\neq 0
\end{equation}%
{\rm which implies that \cite{Far}\cite{McNeill} }%
\begin{equation}
\langle \Phi ^{3}\rangle =u.
\end{equation}%
{\rm This state is a mimimum of the effective potential. On writing }%
\begin{equation}
R_{i\mu }=\cos \left( \mid \vec{B}_{i\mu }\mid \right) +i\vec{\sigma}.\vec{B}%
_{i\mu }\sin \left( \mid \vec{B}_{i\mu }\mid \right) /\mid \vec{B}_{i\mu
}\mid
\end{equation}%
{\rm we find that }%
\begin{equation}
\log I_{0}^{tr}\left( 2\sqrt{y_{i\mu }}\right) \thicksim {\cal M}%
_{B}^{2}\left( \left( B_{i\mu }^{1}\right) ^{2}+\left( B_{i\mu }^{2}\right)
^{2}\right) +\,{\sl interaction\,\,terms}
\end{equation}%
{\rm with }$M_{B}^{2}=\kappa ^{2}u^{2}${\rm . Moreover a {\it %
parity-invariant }mass is also generated for the fermions by the
strongly-coupled }$U_{S}\left( 1\right) ${\rm \ interactions in }$G${\rm ,
entailing a gap for charge excitations (consistent with the pseudogap
phase). The }$SU\left( 2\right) ${\rm \ group is broken down to }$U\left(
1\right) ${\rm \ and there remains only one massless vector boson, }$B_{\mu
}^{3}${\rm , the dual photon. An important aspect of this }$U\left( 1\right) 
${\rm \ symmetry is that it is {\it compact} and allows for monopole
solutions which are interpreted as instantons in }$2+1$ {\rm dimensions.
This will lead to a complete breaking of the symmetry of the dual theory and
also a small non-perturbative mass for the dual photon consistent with the
pseudogap phase. On taking a naive continuum limit, a low-energy gauge
theory results, which will be the starting point of our analysis. For a
nodal liquid there should be a Lorentz invariant ground state and so it is
to be expected that only the scalar multiplet }$\Phi _{a},${\rm \ }$%
a=1,\cdots ,3${\rm \ can have non-zero expectation values. Lorentz
invariance requires that the ground state expectation values of }$\digamma
_{a,\mu }${\rm \ should vanish. In addition the Vafa-Witten theorem \ \cite%
{Vafa} rules against the formation of parity violating states (i.e. non-zero
values of condensates of }$S_{4}${\rm ). For the properties of the nodal
liquid we shall assume, for these reasons, that }${\cal A}_{\mu },{\cal S}%
_{4}${\rm \ and }${\cal S}_{4\mu }${\rm \ do not appear in the effective
theory. They will be important sufficiently far from the nodes. The fields
that will appear in our continuum theory will be }$A_{a\mu },\Phi _{a}${\rm %
\ and }$z_{a}${\rm \ in such a way that }$SU\left( 2\right) $ {\rm gauge
symmetry is manifest. Higher order couplings, from power counting, are
expected to be less dominant for small momenta, and consequently should not
be important for an effective low energy theory. The }$U_{{\em s}}\left(
1\right) ${\rm \ field will not explicitly appear but its fundamental effect
of spontaneous breaking will be incorporated through a Higgs potential term.
Similarly the fundamental constraint in the theory, because our formulation
is for low doping, can be subsumed into another Higgs potential term with
suitably chosen coefficients to implement the constraint as tightly as
required. Coalescing all these considerations allows us to define our
starting continuum effective lagrangian \cite{Kov}:}

\begin{equation}
{\cal L}_{M}=-\frac{1}{4}F_{\mu \nu }^{a}F^{a\,\mu \nu }+\frac{1}{2}\left(
D_{\mu }^{\left( 1\right) ab}\varphi ^{b}\right) ^{2}-\mu ^{2}%
\overrightarrow{\varphi }^{2}-\lambda \left( \overrightarrow{\varphi }%
^{2}\right) ^{2}+\frac{1}{2}\left( D_{\mu }^{\left( \frac{1}{2}\right)
ab}z^{b}\right) ^{2}+\xi \left( \vec{z}^{2}-1\right) ^{2}  \label{NAEL}
\end{equation}%
{\rm where}$\;D_{\mu }^{\left( 1\right) ab}=\partial _{\mu }\delta
_{ab}+e\varepsilon ^{cab}A_{\mu }^{c},\;D_{\mu }^{\left( \frac{1}{2}\right)
ab}=\partial _{\mu }\delta _{ab}-\frac{i}{2}eA_{\mu }^{c}\sigma _{ab}^{c}$%
{\rm . This theory involves fields in both adjoint and fundamental
representations of }$SU\left( 2\right) $. The former is due to the charge
degrees of freedom and the latter is due to the spin degrees of freedom.This
will be important for our future considerations.{\rm The above lagrangian
captures the qualitative features of our approach, i.e. the symmetries and
the mechanism for their breaking. For }$\mu ^{2}<0${\rm \ the intuitive
classical analysis indicates spontaneous symmetry breaking, and, for }$\mu
^{2}>0$, {\rm full symmetry. A rigorous analysis indicates no breakdown of
analyticity in going from positive to negative }$\mu ^{2}${\rm . Because of
symmetry-breaking, {\it compact} }$U\left( 1\right) $ {\rm gauge theory
emerges out of an explicitly} $SU\left( 2\right) $ {\rm symmetric theory,
and this is also necessary in the gauge formulation of the }$CP^{1}$ {\rm %
model}\ {\rm which} {\rm requires a }$U\left( 1\right) ${\rm \ gauge field %
\cite{CP}. This emergence of a compact }$U\left( 1\right) ${\rm \ is an
important difference from the earlier work of Dorey and Mavromatos \cite{Dor}%
.}

\section{Duality}

{\rm We will first consider the }$z${\rm \ fields to be very massive (i.e. a
large spin gap) and so for the time-being ignore them. The spin gap
disappears where superconductivity starts up \cite{super} and so, we would
expect that as we approach the superconducting transition, our arguments
will need to be modified. The dynamical symmetry breaking crucial to our
analysis is induced by the holon condensate }$\Phi ${\rm \ fields. By
working in the unitary gauge }$\langle \varphi _{3}\rangle \neq 0${\rm ,}$\;$%
{\rm the massless gauge field is }$A_{\mu }^{3}${\rm \ which we will call
the statistical photon (as opposed to the electromagnetic photon). The
remainder of the gauge fields have a mass }$m_{W}$ {\rm and the Higgs field
has a mass }$m_{H}${\rm . These conclusions are based on a perturbative
analysis. However there is also a crucial non-perturbative effect which is
due to instantons \cite{Polyakov} which gives the statistical photon a mass }%
$m_{A}${\rm .The theory in the absence of the }$z$ fields{\rm \ is the
well-known Georgi-Glashow model which has been a toy model for the study of
confinement based on }$Z_{N}$ vortices{\rm . However in our case it has
quite a different interpretation and, for the group }$SU\left( 2\right) $,
is {\rm not a toy model. We will identify some of its non-perturbative
features in terms of the pseudogap phase of HTS. The addition of the }$z$
degrees of freedom changes a discrete global symmetry to a local one \cite%
{Fos}.{\rm \ Since we are considering zero temperature, fields are defined
over }$3${\rm -dimensional space. Consequently the monopoles of the
euclidean }$3${\rm -dimensional version of the theory can be interpreted as
instantons in one lower spatial dimension. These monopoles will affect the
disorder fields used in discussions of duality. At its root duality is an
elaboration \cite{StatDual} of the one due to Kramers and Wannier \cite%
{Kramers} \ for the Ising model. A phenomenological study of nodal liquids
using duality,\ linking the }$XY$\ {\rm model and the Landau-Ginzburg model
for a scalar field,} {\rm has been pursued by Balents, Fisher and Nayak\cite%
{Balents} but is\ in quite a different spirit. }

{\rm \bigskip\ }

We will assemble a dual description bit by bit. {\rm If we ignore the
effects of monopoles, absent in a non-compact }$U\left( 1\right) $ theory,%
{\rm \ then} {\rm our starting theory, represented by a lagrangian }${\cal L}%
_{AH}${\rm ,\ is the Abelian-Higgs (}$AH${\rm ) model. The associated
hamiltonian density }${\cal H}${\rm ,\ in the temporal gauge, takes the form:%
}%
\begin{equation}
{\cal H}=\frac{1}{2}\left( B^{2}+E_{i}^{2}\right) +\pi ^{\ast }\pi +\left|
D_{i}\phi \right| ^{2}-\mu ^{2}\left| \phi \right| ^{2}+\lambda \left| \phi
\right| ^{4}  \label{basicmodel}
\end{equation}%
{\rm where }$\mu ${\rm \ and }$\lambda ${\rm \ are constants, }$D_{i}\phi
=\left( \partial _{i}+ieA_{i}\right) \phi ,\;i=1,2${\rm \ and }$\pi
=\partial _{0}\phi ${\rm , }$\;\pi ^{\dagger }=\partial _{0}\phi ^{\dagger }$%
{\rm ; the non-trivial commutators are }$\left[ A_{j}\left( t,%
\overrightarrow{x^{/}}\right) ,E_{i}\left( t,\overrightarrow{x}\right) %
\right] =\delta _{ij}\delta \left( \overrightarrow{x}-\overrightarrow{x^{/}}%
\right) ${\rm \ and }$\left[ \phi \left( t,\overrightarrow{x^{/}}\right)
,\pi \left( t,\overrightarrow{x}\right) \right] =\delta \left( 
\overrightarrow{x}-\overrightarrow{x^{/}}\right) ${\rm . This model has
classical vortex solutions which are given asymptotically as }%
\begin{eqnarray}
&&\phi \left( t,\overrightarrow{x}\right) 
\mathrel{\mathop{\longrightarrow }\limits_{\left| \overrightarrow{x}\right| \longrightarrow \infty }}%
\phi _{0}e^{i\arg \left( \overrightarrow{x}\right) } \\
&&A_{i}\left( t,\overrightarrow{x}\right) 
\mathrel{\mathop{\longrightarrow }\limits_{\left| \overrightarrow{x}\right| \longrightarrow \infty }}%
-\frac{1}{e}\partial _{i}\arg \left( \overrightarrow{x}\right)
\end{eqnarray}%
{\rm where }$arg\left( \vec{x}\right) ${\rm \ is the angle between }$\vec{x}$%
{\rm \ and one of the co-ordinate axes. The quantum version of the vortex is
represented through the vortex creation operator \cite{origin}\cite%
{Mandelstam}, the disorder variable }$V$,{\rm \ which is dual to }$A_{i\text{
}}${\rm \cite{Marino}. Its commutation relations with }$A_{i\text{ }}${\rm %
and} $\ \phi $ {\rm are }%
\begin{equation}
V\left( t,\overrightarrow{x}\right) A_{i}\left( t,\overrightarrow{y}\right)
=\left( A_{i}\left( t,\overrightarrow{y}\right) -\frac{1}{e}\partial
_{i}^{y}\arg \left( \overrightarrow{y}-\overrightarrow{x}\right) \right)
V\left( t,\overrightarrow{x}\right)  \label{CR2}
\end{equation}

\bigskip {\rm and}%
\begin{equation}
V^{-1}\left( \vec{y}\right) \phi \left( \vec{x}\right) V\left( \vec{y}%
\right) =e^{i\arg \left( \vec{x}-\vec{y}\right) }\phi \left( \vec{x}\right) .
\label{CR1}
\end{equation}

The equal-time commutation relations (\ref{CR2}) and (\ref{CR1}) can be
satisfied with the following representation of $V$%
\begin{equation}
V\left( \overrightarrow{x}\right) =\exp \left( \frac{i}{e}\int d^{2}y\left[
\varepsilon _{ij}\frac{\left( x_{j}-y_{j}\right) }{\left( \overrightarrow{x}-%
\overrightarrow{y}\right) ^{2}}E_{i}\left( \overrightarrow{y}\right) +\arg
\left( \overrightarrow{x}-\overrightarrow{y}\right) j_{0}\left( 
\overrightarrow{y}\right) \right] \right) .  \label{REP}
\end{equation}

{\rm In fact, in the absence of fields in the fundamental representation,
i.e. the }$z$ {\rm fields in the lagrangian, }$V\left( \overrightarrow{x}%
\right) $ ${\rm is}$ {\rm local \cite{Kov},\cite{best}. As the }$z$ {\rm %
fields become lighter and so need to be taken into consideration, }$V$ {\rm %
will become non-local and} {\rm the description has to change in a profound
way. From (\ref{CR2}) it follows that }%
\begin{equation}
\left[ V\left( t,\overrightarrow{x}\right) ,B\left( t,\overrightarrow{y}%
\right) \right] =-\frac{2\pi }{e}V\left( t,\overrightarrow{x}\right) \delta
\left( \vec{x}-\vec{y}\right) =-gV\left( t,\overrightarrow{x}\right) \delta
\left( \vec{x}-\vec{y}\right)   \label{Flux1}
\end{equation}%
and

\begin{equation}
\left[ V\left( t,\overrightarrow{x}\right) ,\Phi \left( t\right) \right]
=-gV\left( t,\overrightarrow{x}\right)  \label{FLUX}
\end{equation}%
{\rm where} $\Phi \left( t\right) =\int d^{2}yB\left( t,\overrightarrow{y}%
\right) $ {\rm , }$B${\rm \ is the magnetic field and }$g$ is the `magnetic'
charge conjugate to $e${\rm . Hence }$V$ {\rm \ can be seen as creating a
fundamental unit }$g$ {\rm of magnetic flux which is compatible with being a
disorder variable. This will be a key observation for us. For the Higg's
phase the flux symmetry is not broken and }$\left\langle V\right\rangle =0$%
{\rm \ whereas in the Coulomb phase }$\left\langle V\right\rangle \neq 0$ 
{\rm \cite{best}. The photon is the Goldstone boson of the broken symmetry.
The phase transition can be interpreted as a condensation of vortices. We
should not get confused here by the mention of phase transitions where
previously we stated that there is no phase transition. There is no
contradiction since we are supressing the }$z$ degrees of freedom. It is the
presence of the fundamental fields which implies the lack of a phase
transition.

{\rm A Landau-Ginzburg approach to the low-energy theory, initiated by 't
Hooft \cite{origin}, leads to a dual lagrangian }${\cal L}_{d}$ which {\rm %
has the form}

\begin{equation}
{\cal L}_{d}=\partial _{\mu }V^{\ast }\partial ^{\mu }V-\Omega \left(
V^{\ast }V\right)  \label{EL}
\end{equation}%
{\rm where }$\Omega ${\rm \ is a Higgs-like potential. The field }$V$ is not
restricted to have strictly unit magnitude since this allows some of the
features of quantum fluctuations to be seen at the classical level.{\rm This
is an effective dual picture valid near the cross-over from Higgs to
confining behaviour ( since a phase transition is absent) . The classical\ 
{\em solitons} in this {\em dual} description are the elementary excitations
in the original quantum theory \cite{origin}\cite{Kov},\cite{Kovner}. Before
we consider the form of the solitons, it is necessary to incorporate the
effect of instantons (i.e. monopoles). The magnetic current }$\widetilde{F}%
^{\mu }$ {\rm is no longer conserved but satisfies}

{\rm \ }%
\begin{equation}
\partial _{\mu }\widetilde{F}^{\mu }\left( x\right) =\frac{1}{2}\partial
_{\mu }\varepsilon ^{\mu \nu \sigma }F_{\nu \sigma }\left( x\right) =k\left(
x\right)
\end{equation}%
{\rm where }$k\left( x\right) ${\rm \ is the non-zero magnetic charge
density ( in analogy to the }$U\left( 1\right) ${\rm \ charge current }$%
j^{\mu }${\rm \ ). It is given by}

{\rm \ }%
\begin{equation}
k\left( x\right) =\frac{1}{2e}\varepsilon ^{abc}\varepsilon ^{\mu \nu \sigma
}\partial _{\mu }\widehat{\varphi }^{a}\left( x\right) \partial _{\nu }%
\widehat{\varphi }^{b}\left( x\right) \partial _{\sigma }\widehat{\varphi }%
^{c}\left( x\right)
\end{equation}%
{\rm and } ${\rm \cite{'tHooft}}$ 
\begin{equation}
F^{\mu \nu }\equiv F_{\mu \nu }^{a}\widehat{\varphi }^{a}-\frac{1}{e}%
\varepsilon ^{abc}\widehat{\varphi }^{a}D_{\mu }\widehat{\varphi }^{b}D_{\nu
}\widehat{\varphi }^{c}
\end{equation}%
{\rm where }$\widehat{\varphi }^{a}=\frac{\varphi ^{a}}{\sqrt{\varphi
^{b}\varphi ^{b}}}${\rm . Consequently, when there exist monopole solutions,
the magnetic flux is no longer conserved. This complication results in a
form for }$V$ {\rm which is identical to that already given with }$%
E_{i}\left( \overrightarrow{y}\right) $ {\rm \ replaced by} $F^{0i}\left( 
\overrightarrow{y}\right) $. {\rm For a monopole centred at the origin of }$%
3 ${\rm -dimensional euclidean space }%
\begin{equation}
\partial _{\mu }\widetilde{F}_{\mu }=\frac{4\pi }{e}\delta ^{3}\left(
x\right) .  \label{MON}
\end{equation}%
{\rm On integrating over }$\left[ t_{0},t_{0}^{^{\prime }}\right] \times
R^{2}$ {\rm and remembering that there is a mass gap we conclude from (\ref%
{MON}) that}

\begin{equation}
\Phi \left( t_{0}\right) -\Phi \left( t_{0}^{^{^{\prime }}}\right) =\frac{%
4\pi }{e}.  \label{FluxMON}
\end{equation}%
{\rm This explicitly shows that the flux is not conserved. In the presence
of a finite number of monopoles and anti-monopoles, the flux can change with
time by a multiple of }$\frac{4\pi }{e}${\rm . The unitary operator }$%
U_{\alpha }\left( t\right) ${\rm \ for a flux transformation is given in
general by }%
\begin{equation}
U_{\alpha }\left( t\right) =e^{ie\alpha \Phi \left( t\right) }
\end{equation}%
{\rm but }$\alpha $ {\rm is restricted. In fact for instanton configurations
with Kronecker index }$n,${\rm \ the operators }%
\begin{equation}
U_{n}=e^{i\frac{e}{2}n\Phi }
\end{equation}%
{\rm labelled by an integer }$n${\rm \ is an invariance of the Hilbert space
provided }%
\begin{equation}
U_{n}\left( t_{0}\right) U_{n}^{-1}\left( t_{0}^{^{\prime }}\right) =e^{2\pi
ink}=1
\end{equation}%
{\rm where} $k$ {\rm is an integer. Moreover the commutator (\ref{FLUX})
still holds (with the appropriate forms of the operators) \cite{Kov}. From (%
\ref{FLUX}) the values of the flux on physical states can only be multiples
of }$\frac{2\pi }{e}${\rm . Hence only the operators }$U_{n}${\rm \ with }$%
n=0${\rm \ and }$n=1${\rm \ are independent. Consequently because of
instantons the }$U\left( 1\right) $ {\rm flux symmetry reduces to a }$Z_{2}$%
{\rm \ symmetry and so }${\cal L}_{d}$ {\rm has to have non-}$U\left(
1\right) $ {\rm symmetric terms which nontheless are }$Z_{2}$ {\rm %
symmetric. } {\rm \ Hence }${\cal L}_{d}$ {\rm becomes \cite{origin} }%
\begin{equation}
{\cal L}_{d}=\partial _{\mu }V^{\ast }\partial _{\mu }V-\widetilde{\lambda }%
\left( V^{\ast }V-\widetilde{\mu }^{2}\right) ^{2}-\frac{h^{2}}{4}\left(
V^{\ast 2}+V^{2}\right) +\zeta \left( \varepsilon _{\mu \nu \lambda
}\,\partial _{\nu }V^{\ast }\partial _{\lambda }V\right) ^{2}.  \label{SYMB}
\end{equation}%
{\rm The four derivative term in }${\cal L}_{d}$ is important for describing
the breaking of the adjoint string which is a higher energy phenomena and so
is dominant at short distances. {\rm The system has a mass gap since the
phase of }$V${\rm \ develops a mass }$2\sqrt{h}${\rm . This is interpreted
as the photon acquiring a mass because of instanton effects and is connected
with the absence of long range phase coherence in the pseudogap phase \cite%
{Far}. However there is no genuine phase transition between the confining
and Higgs `phases', and so }${\cal L}_{d}$ {\rm which is most easily derived
through semi-classical methods in the Higgs phase (weak coupling) is valid
also in the confining phase. The difference lies in the spectral properties
i.e. whether }$m_{A}\ll m_{H}$ \ ( in the Higgs phase) or vice versa (in the
confining phase).{\rm \ The mass }$m_{A}$ is {\rm associated with the phase
of }$V${\rm \ }${\rm and}${\rm \ the mass }$m_{H}$ {\rm with the radial part
of }$V${\rm . From semi-classical and weak coupling calculations the
parameters in }${\cal L}_{d}$ {\rm can be identified\ in terms of the masses
and couplings of the microscopic lagrangian }${\cal L}_{M}$ {\rm as follows:}

\begin{eqnarray}
\widetilde{\mu }^{2} &=&\frac{e^{2}}{8\pi ^{2}}  \nonumber \\
\widetilde{\lambda } &=&\frac{2\pi ^{2}m_{H}^{2}}{e^{2}}  \nonumber \\
h &=&m_{A}  \nonumber \\
\zeta &\propto &\frac{M_{W}}{e^{4}m_{H}^{2}}.  \label{wc4}
\end{eqnarray}

\section{Adjoint Texture}

{\rm There are various topological textures which are possible in the dual
theory\cite{Kov},\cite{Kovner}.Quantum features in the original theory can
be seen at a classical level in terms of these textures.} The winding number
of the $V$ field in the texture is given by the charge of the $U\left(
1\right) $ symmetry in ${\cal L}_{M}$. This is encoded in the expression for
the $U\left( 1\right) $ current: 
\begin{equation}
J_{\mu }=-i\frac{2\pi }{e}\varepsilon _{\mu \nu \rho }\partial ^{\nu }\left(
V^{\ast }\overleftrightarrow{\partial }^{\rho }V\right) .
\end{equation}%
Clearly the charge is non-vanishing only where the phase of $V$ is singular.

{\rm An isolated vortex defect has an energy which diverges linearly with
the size of the system.The\ hedgehog configuration is quadratically
divergent and is not energetically favourable owing to the breakdown of the }%
$U\left( 1\right) $ {\rm symmetry to }$Z_{2}$ (due to monopole effects in
the Higgs phase){\rm . In the confining phase the same symmetry picture
arises due to infrared singularities \cite{Cornwall} which is to be expected
if there is no genuine phase transition. This defect is the adjoint string %
\cite{Kovner} and corresponds to the charged }$SU\left( 2\right) $ {\rm %
gauge particles. A finite string and also\ finite energy would arise from a }%
$Z_{2\text{ }}$vortex and anti-vortex. The particles that these dual
structures represent are bound states of the statistical gauge field. A\
statistical gauge field cannot be seen directly but it will leave its
imprint\ on the properties of the holons and spinons. We will describe this
in the remainder of this article. {\rm Its thickness can be estimated to be
inversely proportional to }$m_{A}$\cite{Kov}. In the confinement phase this
is small and the string is narrow. Let us recall that $e^{2}\sim J$ and so
we are in the strong coupling situation for low doping. We contend that the
adjoint string is the basis of stripe behaviour. Firstly let us discuss how
long this might be. Typically stripes are long and thin. We can write $V$ as 
$\rho e^{i\theta }$. For the string to end it needs a vortex structure of $V$
at which to terminate. The size of a vortex core is the region for which $%
\rho $ differs from its vacuum value. Moreover at short distances the
dominant energy term is $\zeta \left( \varepsilon _{\mu \nu \lambda
}\partial _{\nu }V^{\ast }\partial _{\lambda }V\right) ^{2}$. Now the size
of the core is $O\left( \frac{1}{m_{H}}\right) $and so the scale of the core
energy ${\cal E}$ can be estimated by $\zeta m_{H}^{4}\mu ^{4}$. For weak
coupling we can get more information since we can identify the dual
couplings with those from the microscopic theory ( cf (\ref{wc4})). Although
we are really interested in strong coupling, the absence of phase
transitions emboldens us to use weak coupling and extrapolate. We will not
be accurate but the hope is that the qualitative inferences will hold for
strong coupling. Hence we have that the core energy density ${\cal E}$
satisfies ${\cal E}\sim \zeta m_{H}^{4}e^{4}$. It is known that the adjoint
string tension $\sigma _{A}\sim e^{2}m_{A}$ \cite{Kov} and so the length of
the stripe $L$ before it snaps, i.e. the typical length of a stripe, is
given by 
\begin{equation}
\sigma _{A}L\sim \zeta m_{H}^{4}e^{4}\left( \frac{1}{m_{H}}\right) ^{2}.
\end{equation}%
This gives $L\sim \frac{M_{W}}{e^{2}m_{A}}.$For thin stripes it is necessary
that $L\gg \frac{1}{m_{A}}$. It may not be possible to establish without a
detailed numerical analysis. However given the absence of a true phase
transition (in the full theory) we might be able to glean something from an
extrapolation of the weak coupling analysis. From the semi-classical
analysis of Polyakov~\cite{Polyakov} 
\begin{equation}
m_{A}^{2}\sim \frac{16\pi ^{2}}{e^{2}}\frac{m_{W}^{\frac{7}{2}}}{e}\exp %
\left[ -\frac{2\pi m_{W}}{e^{2}}\epsilon \left( \frac{m_{H}}{m_{W}}\right) %
\right]
\end{equation}%
where $\epsilon $ is a slowly varying function but limited in range to $%
1<\epsilon <1.787$. This exponential factor when substitute into the
estimate for $L$ gives an indication that for strong coupling the ratio of
the length to the width can be a large factor. Owing to the uncertainties of
the extrapolation we cannot be more quantitiative. In the strong coupling
limit with $m_{A}\geq m_{H}$ the width of the stripe is inversely
proportional to $m_{A}$. In the absence of the pseudogap the width of the
stripe goes to infinity since $m_{A}\rightarrow 0$ and so the stripe
structure disappears. These one dimensional structures which are related to
the adjoint field lead to one-dimensional charge structures.These will play
a catalytic role for the formation of domain structures. There are however
other crucial features which need to be incorporated within this picture.The
first is that holon transport will be along the stripe. The second is the
antiphase nature of the spin structure where the staggered magnetisation
will point in opposite directions on either side of the stripe. In order to
investigate these aspects it will be necessary to consider the spinon
degrees of freedom which are in the fundamental representation.

\section{Scalar fundamental}

The model of (\ref{NAEL}) has additional terms in the scalar field $z$ which
transforms as the fundamental representation of $SU\left( 2\right) $. This
actually is necessary for establishing that there is an analytic path in
coupling space which connects the Higgs and Coulomb `phases' \cite{Frad}. We
have relied on this result in the previous section even though the $z$%
-degrees of freedom did not appear directly.The dynamical symmetry breaking
of $SU\left( 2\right) $ to $U\left( 1\right) $ allows the $CP_{1}$ model for
the $z$ s which is necessary for a description of the magnetic degrees of
freedom. {\rm The }$L${\rm \ of \ref{NAEL} has a global }$U\left( 1\right) $%
{\rm \ symmetry given by }%
\begin{equation}
z^{a}\rightarrow e^{i\alpha }z^{a}  \label{MAGNETIC}
\end{equation}%
{\rm and the associated quantum number will be called a `magnetic' number
since this symmetry is just part of our global rotation symmetry (\ref{ROTNL}%
). In our effective dual lagrangian this global }$U\left( 1\right) ${\rm \
will play an important role.}

{\rm The previously introduced vortex field }$V${\rm \ is neutral under this
transformation and is associated with the degree of freedom derived from the
frustration of hole motion in an antiferromagnetic background. In the
prescence of the }$z$'s $V$ is non-local \cite{Fos}.This is related to the
fact that the Wilson loop\ has a perimeter law independent of confinment or
the 
\mbox{$\vert$}%
Higgs phase and so $V$ could not be a local order parameter. Once $V$ is
allowed to be a local order parameter it is possible to give a standard and
heuristic argument by splitting the minimal surface inside the loop into
little areas; an area or perimeter law that differentiates between
confinement and the Higgs \ phase ensues.We will find that there are {\rm %
new solitons for }$V${\rm , the fundamental string, in the presence of these
magnetic degrees of freedom. At energy scales where magnetic numbers appear,
in the dual representation it is necessary to add a magnetically charged
complex field }$U$ to incorporate this phase symmetry{\rm . In our effective
lagrangian the field will be important for inducing spontaneous symmetry
breaking of the }$U\left( 1\right) ${\rm \ gauge group down to }$Z_{2}${\rm %
. Moreover defects (vortices) of the }$U${\rm \ field ( given by the zeros
of }$U${\rm ) will be contained within the core of the }$V${\rm \ defect and
represent spinons. The }$U$ field is thus a dual represenation{\rm \ of the }%
$z$ field.{\rm \ By duality} vortices in $z$ ( i.e. skyrmions) will
represent smooth configurations of $U$ ( or $U$ condensates). {\rm Since in
the pseudogap phase there is a spin gap, it is necessary for the spinon
fields to be massive (with mass }$M_{z}${\rm ). In our formalism this is
seen from the constraint on the }$z${\rm \ fields. In the presence of matter
fields in the fundamental representation, the hamiltonian involving }$V${\rm %
\ fields is not globally }$Z_{2}${\rm \ invariant but has a local }$Z_{2}$%
{\rm \ invariance. Consequently it is necessary to have a lagrangian which
is manifestly }$Z_{2}${\rm \ gauge invariant. The amalgamation of the above
considerations leads to a dual lagrangian }$L_{D}${\rm \ \cite{Fos} with the
following structure:}

\begin{eqnarray}
L_{D} &=&-\frac{1}{4e^{2}}f_{\mu \nu }f^{\mu \upsilon }+\frac{1}{2}\left|
\left( \partial _{\mu }-\frac{i}{2}b_{\mu }\right) V\right| ^{2}+\frac{1}{2}%
\left| \left( \partial _{\mu }-ib_{\mu }\right) U\right| ^{2}-\lambda
_{v}\left( V^{\ast }V-\mu ^{2}\right) ^{2}  \nonumber \\
&&-\lambda _{u}\left( U^{\ast }U-u^{2}\right) ^{2}-\xi \left( V^{2}U^{\ast
}+V^{\ast 2}U\right) +\zeta \left( \varepsilon _{\mu \nu \lambda }\partial
^{\nu }V^{\ast }\partial ^{\lambda }V\right) ^{2}  \label{MainLag}
\end{eqnarray}%
{\rm where }$f_{\mu \nu }=\partial _{\mu }b_{\nu }-\partial _{\nu }b_{\mu }$%
{\rm \ . We consider }$\lambda _{u}\gg \lambda _{v}${\rm \ as well as }$u\gg
\mu ${\rm . The gauge transformations of }$U${\rm \ and }$V${\rm \ are
interlocked because, when }$V\rightarrow Ve^{i\alpha }${\rm , it is
necessary that }$U\rightarrow Ue^{2i\alpha }${\rm (manifest from the term
proportional to }$\xi ${\rm ). This lagrangian is a generalization of that
given earlier (\ref{SYMB}). }Although we cannot provide a rigorous
justification for the effective lagrangian $L_D$ , we will show that it
embodies the important symmetries that are present in ${\cal L}_{M}$ (\ref%
{NAEL}).

\bigskip {\rm \ It will first be shown how in the the absence of fundamental
fields }$L_{D}{\rm \ }${\rm reduces to the lagrangian in (\ref{SYMB}).} We
first note that the following reparametrisation \cite{Fos} of $e,u$ and $z,$
with the assumption that $y,x$ and $\kappa $ are independent of $M_{z}$, 
\begin{equation}
e^{2}=\frac{y}{M_{z}},\;u^{2}=xM_{z}\text{ },\;\zeta =\frac{\kappa }{u}\text{%
{}}  \label{scaling}
\end{equation}%
leads to the desired reduction. In the limit $M_{z}\rightarrow \infty $, $%
e\rightarrow 0$ and the vector boson mass $m_{b},$ which satisfies $%
m_{b}^{2}=e^{2}u^{2},$ remains finite. This results in a lagrangian with
terms which coincide with \ref{SYMB} together with terms involving decoupled 
$\overrightarrow{W}$ fields. The above parametrisation can be deduced for
large $M_{z}$ with $y\propto \Lambda ^{2}$ as will be shown below.

\bigskip

The lagrangian $L_D$ has two global $U\left( 1\right) $ symmetries. The
associated currents are

\begin{equation}
{\em j}^{\mu }=\frac{1}{e^{2}}\partial _{\nu }f^{\nu \mu }  \label{current1}
\end{equation}

and 
\begin{equation}
\widetilde{f}_{\mu }=\varepsilon _{\mu \nu \lambda }f^{\nu \lambda }.
\label{current2}
\end{equation}

In terms of ${\cal L}_{M}$ we can make the following identification:%
\begin{equation}
\frac{1}{2\pi }\widetilde{f}_{\mu }=j_{\mu }^{M}
\end{equation}

\bigskip where the current $j_{\mu }^{M}$ connected with the transformation %
\ref{MAGNETIC} is 
\begin{equation}
j_{\mu }^{M}=i\left( z^{a^{\ast }}\overleftrightarrow{\partial }_{\mu
}z^{a}\right) .
\end{equation}

Also the charge $Q=\int $ ${\em j}^{0}d^{2}x$ \ can be expressed as 
\begin{equation}
Q=\int d^{2}x\partial _{i}\left( \frac{1}{e^{2}}\varepsilon _{ij}\widetilde{f%
}_{j}\right) 
\end{equation}

and so is related to the vorticity of the magnetic current. We will now
demonstrate that this vorticity of the magnetic current is created by the
operator $V$. Using Gauss's law in 2-dimensions, we can rewrite the operator
(\ref{REP}) as the exponential of a line integral over a path $C_{x}$which
is the branch cut of the multi-valued function $arg\left( y-x\right) $ at $x$%
. For clarity we will denote this form of the operator by $V_{C_{x}}$($=\exp
\left( \frac{2\pi i}{e}\int_{C_{x}}ds_{i}\varepsilon _{ij}E_{j}\right) $).
This operator is identified with $V$ in $L_{D}$ after gauge fixing. It can
then be shown that \cite{Fos}

\begin{equation}
V_{C_{x}}^{\dagger }j_{i}^{M}\left( x\right) V_{C_{x}}=j_{i}^{M}\left(
x\right) +\pi n_{i}^{C}\left( x\right) \delta \left( x\in C\right)
z^{a^{\ast }}\left( x\right) z^{a}\left( x\right)   \label{MagneticCurrent}
\end{equation}

where $n_{i}^{C}\left( x\right) $ is the normal to $C_{x}$ at $x$. The
vorticity density $\omega $ of the magnetic current is defined as 
\begin{equation}
\omega =i\varepsilon _{ij}\partial _{i}\left( \frac{z^{a^{\ast }}%
\overleftrightarrow{\partial }_{j}z^{a}}{z^{b^{\ast }}z^{b}}\right) .
\label{current3}
\end{equation}

{}From arguments similar to that used in (\ref{MagneticCurrent}) 
\begin{equation}
V_{C_{x}}^{\dagger }\omega \left( y\right) V_{C_{x}}=\omega \left( y\right)
+\pi \delta ^{\left( 2\right) }\left( x-y\right)
\end{equation}

and so $V_{C_{x}}$ creates vorticity $\pi $, i.e. half a unit of vorticity.
{}From the gauge couplings of $U$ and $V$ in $L$ it is clear that $U$ will
create twice the vorticity produced by $V$.

This derivation allows us to identify $e^{2}$ in $L_{D}$. Indeed, from the
relations (\ref{current1}) and (\ref{current2}) we can deduce that 
\begin{equation}
\omega =\varepsilon _{ij}\partial _{i}\left( \frac{2\pi }{e^{2}}%
j_{j}^{M}\right) 
\end{equation}

and so $e^{2}\propto z^{b^{\ast }}z^{b}$. In the effective theory it is a
good approximation to replace  $z^{b^{\ast }}z^{b}$ by its expectation value
and so $e^{2}\propto \left\langle z^{b^{\ast }}z^{b}\right\rangle .$

This identification allows us to verify the scaling of $e$ in (\ref{scaling}%
). Indeed, after performing the frequency integration, the leading estimate
for $\left\langle z^{b^{\ast }}z^{b}\right\rangle $ is 
\begin{equation}
\left\langle z^{b^{\ast }}z^{b}\right\rangle =\int d^{2}p\,\frac{1}{\left(
p^{2}+M_{z}^{2}\right) ^{\frac{1}{2}}}\propto \frac{\Lambda ^{2}}{M_{z}}
\end{equation}

which allows us to infer that $y\propto \Lambda ^{2}$ where $\Lambda $ is a
suitable high momentum cut-off that defines the effective theory and so
verify (\ref{scaling}).

We now need to examine the influence of the massive spinon fields. For
energies much less than the scale $u^{2}$ we saw that the $U$ degree of
freedom was frozen and that ${\cal L}_{d}$ (\ref{SYMB}) was recovered which
led to the adjoint string. However there are other configurations which we
have so far neglected and are crucial to the development of the domain wall
structure central to our considerations. For these there are points where $U$
vanishes and so the unitary gauge transforming $U$ to $u$ cannot be fixed.
The vortex of $U$ carries \bigskip magnetic quantum number. From $L_{D}$, by
examining the covariant derivative structure, it is readily seen that for
finite energy for both $V$ and $U$ vortices it is necessary that a single
winding of the $V$ field is accompanied by a double winding of the $U$
field. Moreover, since sufficiently {\it far} from the $U\ $vortex a unitary
gauge form of $U$ is valid, the field $V$ forms a one-dimensional adjoint
string . In the absence of $U$ vortices it must be accompanied by an
anti-defect which, for $Z_{2}$, is the same as the defect itself. This is
the incipient charge order which, on introduction of the $z$ 's will lead to
a special type of spin ordering of the staggered magnetization. Furthermore
charge transport will also be along the texture. Before that it is necessary
to establish the domain wall structure in $V$ in the presence of fundamental
charges, through an argument due to Fosco and Kovner\cite{Fos}. We recall
that the fundamental charges in the dual picture are represented as the
vortices in $U$.

\bigskip Let us see how this happens by using the parametrisation $%
U=ue^{i\theta }$ and $V=\rho e^{i\varphi }$. The dual lagrangian ${\cal L}%
_{D}$ has the form 
\begin{equation}
{\cal L}_{D}=\frac{1}{2}\left( \left( \partial _{\mu }\rho \right) ^{2}+\rho
^{2}\left( \partial _{\mu }\varphi -\frac{1}{2}b_{\mu }\right) ^{2}\right)
-4\zeta \rho ^{2}\left( \varepsilon _{\mu \nu \lambda }\partial ^{\nu }\rho
\partial ^{\lambda }\varphi \right) ^{2}-\lambda _{v}\left( \rho ^{2}-\mu
^{2}\right) ^{2}-2\xi \rho ^{2}u\cos \left( 2\varphi -\theta \right)
\label{duallagr}
\end{equation}%
and at long wavelengths the equation of motion for \ $V$ reduces to

\begin{equation}
\partial ^{2}\rho =\rho \left( \partial _{\mu }\varphi -\frac{1}{2}b_{\mu
}\right) ^{2}-4\lambda _{v}\left( \rho ^{2}-\mu ^{2}\right) \rho -4\xi \rho
u\,\cos \left( 2\varphi -\theta \right)
\end{equation}%
and 
\begin{equation}
\partial _{\mu }\left( \rho ^{2}\left( \partial _{\mu }\varphi -\frac{1}{2}%
b_{\mu }\right) \right) =4\xi \rho ^{2}u\sin \left( 2\varphi -\theta \right)
.
\end{equation}

\bigskip

{}From (\ref{duallagr}) it is clear that winding numbers in $U$ and $b_{\mu }
$ \ are in correspondence while $V$ winds only half as much, i.e. far from a
vortex $\varphi =\frac{\theta }{2}$. This is the crucial intuitive reason
for the formation of the domain wall for \ $V$. Consider an adjoint string
with vortex and antivortex at its ends and then introduce a spinon at each
end.We would like investigate the effect this has on the nature of the
texture for $V$. Explicitly consider $U$ vortices, the spinons, in the plane
at $\vec{x}=\left( \pm a,0\right) $. Far from the vortices 
\begin{equation}
U\left( \vec{x}\right) =u\exp \left( i\theta \left( \vec{x}\right) \right) 
\end{equation}%
with 
\begin{equation}
\theta \left( \vec{x}\right) =\tan ^{-1}\left( \frac{x_{2}}{x_{1}-a}\right)
+\tan ^{-1}\left( \frac{x_{2}}{x_{1}+a}\right) 
\end{equation}%
due to the winding of the two vortices. As $x_{2}\longrightarrow \infty $, $%
\theta \left( \vec{x}\right) \longrightarrow \pi $ and $\varphi
\longrightarrow \frac{\pi }{2}$. Similarly as $x_{2}\longrightarrow -\infty $%
, $\theta \left( \vec{x}\right) \longrightarrow -\pi $ and $\varphi
\longrightarrow -\frac{\pi }{2}$. Consequently there is a domain wall formed.

We will now discuss the antiphase properties of the staggered magnetiztion
in the microscopic picture ( as opposed to the dual picture). The above
domain wall structure translates into a domain wall structure in the flux of
the statistical field due to (\ref{FLUX}). We need to find solutions of the $%
CP^{1}$ field $z$ in the presence of this domain structure. The
configuration taken up by this field is that of  vortices on one side of the
domain wall and an anti-vortices on the other. We can infer this from {\rm \
the form of a single static }$CP_{1}$ vortex 
\begin{equation}
z_{a}\left( \vec{x}\right) =\left( 
\begin{array}{c}
\cos \frac{f\left( r\right) }{2}\ e^{-\frac{i}{2}\arg \left( \hat{x}\right) }
\\ 
\sin \frac{f\left( r\right) }{2}\ e^{\frac{i}{2}\arg \left( \hat{x}\right) }%
\end{array}%
\right) 
\end{equation}%
which is centred at the origin and where \ $r=\left| \vec{x}\right| ,\ \hat{x%
}=\frac{\vec{x}}{r},\arg \left( \hat{x}\right) =\arctan \left( \frac{x_{2}}{%
x_{1}}\right) $ and $f\left( r\right) =2\arctan \frac{s}{r}$, $s$ being an
arbitrary scale factor which will be the thickness of the domain wall i.e. $%
s\sim \frac{1}{m_{A}}$. ( In terms of the staggered magnetisation $\vec{n}$ (%
$n_{i}=z^{a\,\ast }\left( \sigma _{i}\right) _{ab}z^{b})$ this configuration
is that of a skyrmion viz. $\vec{n}\left( \vec{x}\right) =\left( \sin
f\left( r\right) \,\hat{x},\cos f\left( r\right) \right) $.) The winding
number of the vortex is the flux of the `magnetic' field $B$. The form of $B$
is 
\begin{equation}
B\left( \vec{x}\right) =\frac{s\sin \left( 2\arctan \left( \frac{s}{\sqrt{%
x_{1}^{2}+x_{2}^{2}}}\right) \right) }{\sqrt{x_{1}^{2}+x_{2}^{2}}\left(
x_{1}^{2}+x_{2}^{2}+s^{2}\right) }
\end{equation}%
and so falls off rapidly away from the vortex. Consequently on one side of
the domain wall there is a skyrmion and on the other an anti-skyrmion in
terms of the staggered magnetisation. Because of the lack of rotational
symmetry in the presence of the domain structure, we expect, in keeping with
the symmetry of the domain wall (or stripe), a row of vortices and
anti-vortices on either side of the stripe. This means that there is an
anti-phase structure for the staggered magnetisation.

In the microscopic picture the nodal fermions satisfy a Dirac equation (with
real time) which near the domain wall takes the form 
\begin{equation}
\left( i\not{\partial}-\epsilon _{a}e\not{A}-\epsilon _{a}m\right) \psi
_{a}=0  \label{Dirac}
\end{equation}%
where $\epsilon _{1}=1$ and $\epsilon _{2}=-1$. (Because of the signs
associated with the masses there is no parity breaking in the model \cite%
{Dor}.) A related construction in a non-relativistic theory for staggered
magnetic fields can be found in \cite{Oleg}. We consider a domain wall along
the $x$-axis with $\vec{A}\left( \vec{x}\right) =B\left( 0,\left|
x_{2}\right| ,0\right) $. On \ making the ansatz for $\psi $ of a
propagating solution 
\begin{equation}
\psi _{a}\left( x_{0},x_{1},x_{2}\right) =\exp \left( i\left( \omega
x_{0}-kx_{1}\right) \right) \chi _{a}\left( x_{2}\right) ,  \eqnum{Ansatz}
\end{equation}%
where $\chi _{a}$ is a 2-component spinor for each $a$ we deduce that 
\begin{equation}
\left( \frac{d^{2}}{d\xi _{a}^{2}}-\frac{\xi _{a}^{2}}{4}+\theta -\frac{1}{2}%
\epsilon _{a}\,sgn\left( x_{2}\right) \gamma ^{0}\right) \chi _{a}=0
\end{equation}%
where $\theta =\frac{\omega ^{2}-m^{2}}{2eB}$ and $\xi _{a}=\sqrt{\frac{2}{eB%
}}\left( k+\epsilon _{a}eB\left| x_{2}\right| \right) $ from (\ref{Dirac}).
Although these equations hold near the stripe, we can check whether there
are propagating solutions localised in $x_{2}$. If so then the stripes could
conduct. In our situation $\xi _{1}\rightarrow \infty $ and $\xi
_{2}\rightarrow -\infty $ when $\left| x_{2}\right| \rightarrow \infty $.
For $a=1$ we impose the boundary conditions at $\xi _{1}\rightarrow \infty $
and $y=0;$ for $a=2$ we impose the boundary conditions at $\xi
_{2}\rightarrow -\infty $ and $y=0$. The symmetric $\left( +\right) $ and
antisymmetric $\left( -\right) $ solutions can be written as%
\begin{equation}
\chi _{1}(y)=\left( 
\begin{array}{c}
D_{\theta -1}(\xi _{1}) \\ 
D_{\theta }(\xi _{1})%
\end{array}%
\right) ~~\;~y>0,~\;~\chi _{1}(y)=\pm \left( 
\begin{array}{c}
D_{\theta }(\xi _{1}) \\ 
D_{\theta -1}(\xi _{1})%
\end{array}%
\right) ~~\;~y<0,  \label{sol1}
\end{equation}

\noindent in terms of parabolic cylinder functions $D_{\theta }$ and
similarly the solutions vanishing at $\xi _{2}=-\infty $ are

\begin{equation}
\chi _{2}(y)=\left( 
\begin{array}{c}
D_{\theta }(-\xi _{2}) \\ 
D_{\theta -1}(-\xi _{2})%
\end{array}%
\right) ~~\;~~y>0,~~~~\;~~\chi _{2}(y)=\pm \left( 
\begin{array}{c}
D_{\theta -1}(-\xi _{2}) \\ 
D_{\theta }(-\xi _{2})%
\end{array}%
\right) ~~\;~y<0.  \label{sol2}
\end{equation}

\noindent We are still left with one boundary condition at $y=0$: for each
sublattice $a$, we impose the condition

\begin{equation}
\lim_{y\to 0^+}\chi_a(y)=\lim_{y\to 0^-}\chi_a(y),
\end{equation}

\noindent such that we have for both sublattices

\begin{equation}
D_{\theta -1}(\epsilon _{a}\xi _{0})=\pm D_{\theta }(\epsilon _{a}\xi
_{0})~,~~~\;~\xi _{0}=k\sqrt{\frac{2}{eB}},
\end{equation}%
\noindent which gives us the dispersion relation $\theta (\xi _{0})$ or
equivalently $\omega (k)$. It is possible to solve these relations which are
consistent (and in the non-relativistic limit give rise to modified Landau
levels). The leading asymptotic behaviour 
\begin{equation}
D_{p}\left( z\right) \sim e^{-\frac{^{z^{2}}}{4}}z^{p}\;,\;\left| z\right|
\rightarrow \infty \text{ and}\;\left| \arg z\right| <\frac{3}{4}\pi
\end{equation}%
and so the wavefunctions are localised on the scale of the `magnetic' length
and so localised conduction along the stripe is consistent within our
framework.

\bigskip

\bigskip

\section{Discussions}

In this work we have started from a $SU(2)\otimes U_{s}(1)$ gauge field
theory of spin-charge separation, based on an approximate particle-hole
symmetric formulation upon doping lightly from half-filling. The strongly
coupled $U_{s}(1)$ group is a symmetry only of the fermionic sector,
expressing appropriate frustrations of holons. Once $U_{s}(1)$ is integrated
out in the path integral one arrives at an effective lagrangian with a
broken $SU(2)\rightarrow U(1)$ phase. In this picture, the presence of
doping is responsible for a dynamical breaking of the initial $SU(2)$ spin
symmetry.

Non-peturbative effects, due to the compactness of the $U(1)$ unbroken
subgroup, are responsible for giving a mass in the associated gauge boson
(``statistical photon''). This leads to a pseudogap phase, that is a holon
mass-gap phase without phase coherence. In this article we have discussed in
some detail some properties of a formalism {\it dual} to the above mentioned
effective lagrangian, which proved very useful in shedding light on various
important physical properties of the original theory. In particular, we have
argued in detail how the dual formalism can explain in a natural way the
stripe phase of the underlying microscopic theory. We have associated such
stripes with appropriate topological textures (domain walls) of the dual
lagrangian. The r\^{o}le of vortices of the dual lagrangian as appropriate
dual configurations of the original spinon (magnon $z$) degrees of freedom
has also been pointed out. Moreover, anti-phase properties of the staggered
magnetization in the microscopic picture (rather than the dual one) have
been studied by making the observation that the domain wall structure in the
dual lagrangian corresponds to a domain wall structure in the flux of the
statistical photon.

We would now like to conclude by making a few speculative remarks on the
r\^ole of nodal excitations to the passage from the pseudogap to the
superconducting phase of the underlying antiferromagnetic system. The
important point to realize is that, in the context of our relativistic nodal
theory, superconducitivity arises in the way explained in ~\cite{Dor}\cite%
{Far}, only in the case of a {\it strictly massless} statistical photon,
i.e. a photon that remains massless non-perturbatively. This photon plays
the r\^ole of the massless pole in the current-current correlator, which is
the basic Landau criterion for superconductivity. In this model, the
pseudogap phase -studied in the present paper- can be succeeded by a
superconducting phase if and only if the non-perturbative monopole effects,
responsible for the statistical photon mass, are suppressed.

We have conjectured in \cite{Far,McNeill} that this may happen in the case
where there is a {\it dynamical supersymmetry} in the effective theory
between spinon and holon degrees of freedom. Such a situation has been
argued by Mavromatos and Sarkar~\cite{MavSar} to characterise specific
points in phase space of some extended $t-j$ models under the spin-charge
non-Abelian separation ansatz (\ref{scan}). Subsequently, it was argued by
Alexandre et al~\cite{Alexandre} that the continuum composite field theory
of such supersymmetric theories, obtained after the integration of the
strongly coupled $U_{s}(1)$ group, exhibits extended $N=2$ supersymmetry,
due to the low-dimensionality; this arguably can provide the necessary
mechanism for the masslessness of the statistical photon, and thus a passage
from the pseudogap to the superconducting phase.

It should be remarked that the presence of extended supersymmetries opens up
the way for some exact results in the phase diagram of such systems. An
interesting question is to formulate the duals of such supersymmetric
theories and study their properties in detail along the lines presented
above, e.g. the fate of textures {\it etc}. This is left for future work. It
remains to be seen, of course, whether realistic microscopic models for
antiferromagnets can be constructed which, in some regions of their
parameter space, could exhibit dynamical supersymmetries between spinon and
holon degrees of freedom, capable of explaining phenomenologically the rich
phase diagrams of the high-temperature superconductors observed
experimentally. We believe that the present work, along with those in \cite%
{MavSar}\cite{Alexandre}, constitute useful contributions to this programme.

{\rm \bigskip }

\section*{Acknowledgement}

{\rm We acknowledge funding from the Leverhulme Trust for this research and
useful discussions with Dr. J Alexandre.}

\end{document}